\documentclass[aps]{revtex4-1}

\usepackage{graphicx}
\usepackage{amsmath}
\usepackage{amssymb}
\usepackage{xcolor}

\begin{document}


\title{Planar immersion lens with metasurfaces} 



\author{John S. Ho$^{1}$}
\author{Brynan Qiu$^{2}$}
\author{Yuji Tanabe$^{1}$}
\author{Alexander J. Yeh$^{1}$}
\author{Shanhui Fan$^{1}$}
\author{Ada S. Y. Poon$^{1}$}
\email[]{Electronic mail: adapoon@stanford.edu}

\affiliation{$^{1}$Department of Electrical Engineering, Stanford University, California 94305, USA}
\affiliation{$^{2}$Department of Electrical Engineering, California Institute of Technology, California 91125, USA}


\date{\today}

\begin{abstract}
The solid immersion lens is a powerful optical tool that allows light entering material from air or vacuum to focus to a spot much smaller than the free-space wavelength. Conventionally, however, they rely on semispherical topographies and are non-planar and bulky, which limits their integration in many applications. Recently, there has been considerable interest in using planar structures, referred to as metasurfaces, to construct flat optical components for manipulating light in unusual ways. Here, we propose and demonstrate the concept of a planar immersion lens based on metasurfaces. The resulting planar device, when placed near an interface between air and dielectric material, can focus electromagnetic radiation incident from air to a spot in material smaller than the free-space wavelength. As an experimental demonstration, we fabricate an ultrathin and flexible microwave lens and further show that it achieves wireless energy transfer in material mimicking biological tissue.
\end{abstract}


\maketitle 

\section{Introduction}

\begin{figure}
   \centering
   \includegraphics[width=7cm]{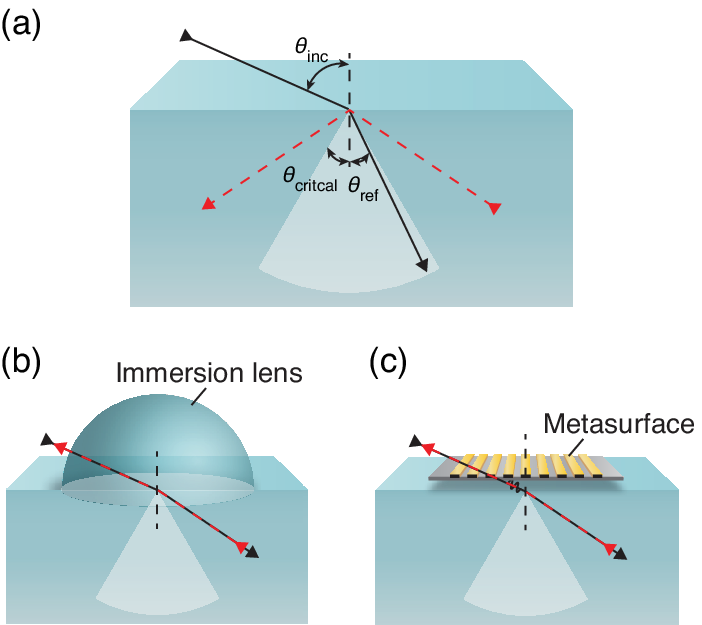} 
   \caption{Refraction at a planar interface between air and high-index material. (a) Ordinary refraction. Light cannot be refracted beyond the critical angle $\theta_{\text{critical}}$ since the corresponding beam is trapped by total internal reflection (dashed red line). (b) Refraction with an immersion lens. The lens enables light to both enter (black line) and escape (dashed red line) material at a forbidden angle. Light can thus be focused to $\sim\lambda/n$, where $n$ is the material refractive index. (c) Refraction with a metasurface. The metasurface changes the amplitude and phase of light, enabling it to refract at a forbidden angle.}
   \label{fig:schematic}
\end{figure}

When light is focused from air into material, refraction at the air-material interface determines the diffraction limit. Conventional optical lenses, placed in the far-field of the interface, control only propagating wave components in air. As a result, their focusing resolution in material is diffraction-limited at the free-space wavelength $\lambda$ because higher wavevector components in material cannot be accessed by far-field light~\cite{Milster_1999}. These high wavevector components correspond to plane waves propagating at angles greater than the critical angle, which are trapped in the material by total internal reflection [Fig.~\ref{fig:schematic}(a)]. 

Solid immersion lenses are semispherical domes of high-index material placed at or near the air-material interface that allow light to access these ``forbidden'' angles of refraction. This capability enables light to be focused to a spot much smaller than the free-space wavelength, with a diffraction-limited resolution set by the material wavelength $\sim\lambda/n$ [Fig.~\ref{fig:schematic}(b)]. Solid immersion lenses have found extensive use in many applications, including imaging~\cite{Mansfield_1990}, data storage~\cite{Terris_1994}, and lithography~\cite{Ghislain_1999}. They are, however, intrinsically three-dimensional and bulky~\cite{Sendur_2005}. Replacing conventional solid immersion lenses with flat counterparts would afford opportunities for integration in complex systems, including nanophotonic chips or, in the low frequency regime, conformal biomedical devices.

In this paper, we propose and demonstrate a planar immersion lens based on metasurfaces. Metasurfaces are flat devices consisting of structured arrays of subwavelength apertures or scatterers that provide an abrupt change in electromagnetic properties as light propagates across the surface~\cite{Yu_2014}. The properties of metasurfaces can be tuned by varying the parameters of the individual subwavelength elements to form a desired spatially varying response. This extraordinary freedom in design has been used to create devices that generate negative refraction~\cite{Yu_2011,Ni_2012}, light vortices~\cite{Genevet_2012}, flat lensing~\cite{Aieta_2012,Chen_2012}, holograms~\cite{Ni_2013,Huang_2013}, and other unusual interface phenomena in both optical~\cite{Kildishev_2013} and microwave~\cite{Li_2012,Sun_2012a} regimes. We report metasurfaces based on electrically thin metallic strips with deep subwavelength spacing that allow radiation incident from air to refract into forbidden angles in material [Fig.~\ref{fig:schematic}(c)]. We then use this capability to develop a thin and planar device that reproduces the functionality of a solid immersion lens. The device is fabricated on a flexible substrate and demonstrated at microwave frequencies.

\section{Refraction at Forbidden Angles}

To allow for interface phenomena different from classical reflection and refraction, a metasurface can be placed at the air-material boundary to break translational symmetry at the interface. When the metasurface imparts a phase with constant gradient $\nabla\Phi$ on incident light, propagation is governed by a generalized form of Snell's law~\cite{Yu_2011}.  The law implies that radiation incident at an angle $\theta_{\text{inc}}$ refracts at a forbidden angle $|\theta_{\text{ref}}|>\theta_{\text{critical}}$ if the phase gradient is sufficiently large $|\nabla\Phi| > k_0 - k_{0}\sin|\theta_{\text{inc}}|$ (see Methods). 

\begin{figure}
   \centering
   \includegraphics[width=8cm]{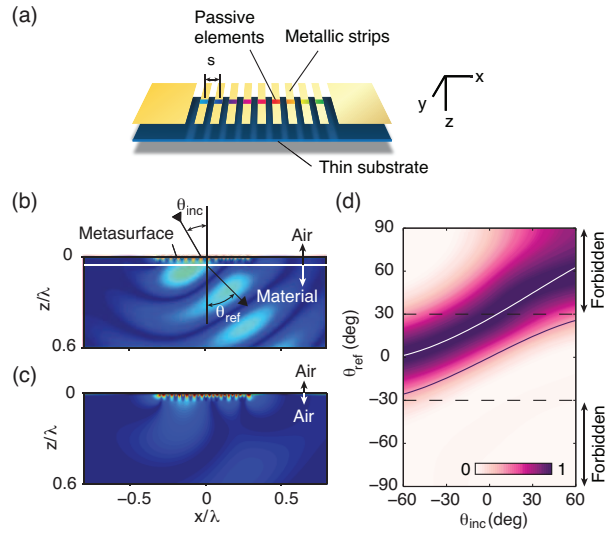} 
   \caption{Refraction at forbidden angles with a metasurface. (a) Metasurface consisting of metallic strips reactively loaded with impedances varying in the $x$ direction. The strips, separated by a distance of $s=\lambda/20$, are periodic in the $y$ direction and impart a phase gradient of $\pi/ 0.55\lambda$ on incident s-polarized radiation. (b) Magnetic field amplitude $|\text{Re}(H_x)|$ below the metasurface in the air gap and material ($n=2$) for a plane wave at $\theta_{\text{inc}}=30^{\circ}$. (c) Magnetic field in air when the material volume is removed. (d) Contour plot of the angular spectrum of the refracted beam as a function of the angle of incidence. The solid white line is the angle of refraction predicted by the generalized Snell's law and the solid purple line ordinary refraction.}
   \label{fig:forbid_refract}
\end{figure}

We implement a phase gradient by non-periodically modulating the surface with subwavelength structures of varying impedances. Fig.~\ref{fig:forbid_refract}(a) shows the metasurface consisting of metallic strips loaded with passive lumped elements (resistors, capacitors, and inductors). At microwave frequencies, these elements can consist of patterned metal traces or commercial impedance components. Across a resonance, the phase of the current in the strips differs from that of the driving electric field by a value between $0$ and $\pi$. By selecting suitable passive elements, taking into account both the intrinsic and mutual impedances of the structures, the spatial phase profile of the transmitted wave can be shaped within this phase range. The use of discrete passive elements considerably simplifies the design as the metasurface can be reconfigured by simply changing the elements~\cite{Grbic_2008,Grbic_2011}. Because coupling is explicitly accounted for in the design, the inter-element spacing can be made deeply subwavelength. The phase range can be extended to the full $0$ to $2\pi$ by exploiting changes in polarization (Berry phase)~\cite{Yu_2011}, incorporating elements with a magnetic response~\cite{Sun_2012b,Pfeiffer_2013}, or cascading multiple layers~\cite{Monticone_2013}, although the limited range achieved here using a single layer is sufficient to realize a planar immersion lens.

Refraction at a forbidden angle is shown in Fig.~\ref{fig:forbid_refract}(b) using the metasurface to create a phase gradient of $\nabla\Phi=\pi/0.55\lambda$ for radiation at 1.5~GHz. The spacing between the elements is $\lambda/20$, and is necessarily subwavelength in order to satisfy sampling requirements. For an s-polarized plane wave incident at $\theta_{\text{inc}}=30^{\circ}$, the beam is entirely refracted to an anomalous angle of $\theta_{\text{ref}}=45^{\circ}$ that lies well beyond the critical angle $\theta_{\text{critical}}=30^{\circ}$. Because the metasurface does not rely on polarization conversion, there is no co-polarized component refracted at an angle dictated by the standard Snell's law~\cite{Sun_2012b,Pfeiffer_2013,Monticone_2013}. Unlike diffraction gratings, which alter the spatial amplitude profile, the metasurface refracts the incident wave by modulating its phase profile and thus does not result in unwanted diffractive orders.

We find that anomalous refraction still occurs when a $\lambda/20$ air gap is introduced between the metasurface and the interface [Fig.~\ref{fig:forbid_refract}(b)]. This effect is closely related to frustrated total internal reflection. In absence of material below the metasurface, the incident wave completely reflects off the metasurface and forms an evanescent wave at the surface [Fig.~\ref{fig:forbid_refract}(c)]. This evanescent wave propagates along the surface in the direction of the phase gradient, which is a non-trivial behavior that cannot be realized with a grating~\cite{Sun_2012b}. When the material is placed in close proximity to the metasurface, the evanescent field phase matches to a propagating wave in the material, allowing the incident beam to tunnel into the material with the net transport of energy across the interface. By varying the angle of incidence relative to the phase gradient, the angular spectrum of the transmitted beam can be made to lie almost entirely in the forbidden region [Fig.~\ref{fig:forbid_refract}(d)]. The results are in agreement with the generalized Snell's law. The spread around the predicted angle is due primarily to finite size of the aperture (see Supplemental Materials).

\section{Optimal Focusing}
\label{sec:optimal_focusing}

In order to design a planar immersion lens, a field source in air that focuses to a $\lambda/n$ spot in material must first be found. Although focusing across a planar interface has been previously studied, classic expressions for the optimal field source consider only far-field light and yield a $\sim\lambda$ focal spot~\cite{Hao_1984a,Wiersma_1996,Wiersma_1997}. We introduced a more general approach in ref.~(21) that accounts for evanescent waves at the interface. We start by formulating an optimization problem over the space of current sheets $\mathbf{j}_{s}$ in the source plane (taken to be $z=0$). The solution to the problem is defined to be the current sheet that maximizes a metric for the degree of focus~\cite{Kim_2013}. For present purposes, we assume that the material is dissipative, allowing small but non-zero loss. We then take the efficiency of work performed on the material as the focusing metric
 \begin{equation}
\eta = \frac{\alpha'' | \mathbf{E}(\mathbf{r}_f)|^2}{ \int d\mathbf{r}~\epsilon''~|\mathbf{E}(\mathbf{r})|^2}
\end{equation}
where $\mathbf{r}_{f}$ is the focal point, $\alpha''$ the imaginary part of the polarizability of the object at the focal point, and $\epsilon''$ the imaginary part of the material dielectric permittivity. We set $\alpha$ to be the polarizability of a ``virtual'' sphere centered at the focal point: the sphere has the same dielectric permittivity as the background material and can be made arbitrarily small (e.g., the diameter of a computational mesh unit). The electric field $\mathbf{E}$ can be found by propagation from the current sheet $\mathbf{j}_s$ as described by the Green's function $\mathbf{G}(\mathbf{r},\mathbf{r}')$. 

The optimization problem can now be considered in the operator formalism. Using Dirac bra-ket notation, we represent $\mathbf{E}$ and $\mathbf{j}_{s}$ respectively as functions $|\psi\rangle$ and $|\phi\rangle$ in Hilbert space. They are related through the operator expression $|\psi \rangle=\hat{G} |\phi \rangle$ where $\hat{G}$ is the Green's function operator. We also define a focal position operator $\hat{\Phi}$ such that the numerator in Eq.~(1) can be written as
\begin{equation}
\langle \psi |\hat{\Phi}| \psi \rangle= \alpha''\int d\mathbf{r}~\delta(\mathbf{r}_f) |\mathbf{E}(\mathbf{r})|^2=\alpha''~|\mathbf{E}(\mathbf{r}_{f})|^2.
\end{equation}
Similarly, a power loss operator $\hat{\Sigma}$ can be defined to yield
\begin{equation}
\langle \psi |\hat{\Sigma}| \psi \rangle=\int d\mathbf{r}~\epsilon''~|\mathbf{E}(\mathbf{r})|^2.
\end{equation}
Optimal focusing occurs when the choice of the source current density $|\phi\rangle$ maximizes Eq.~(1). Focusing can thus be posed as the optimization problem
\begin{equation}
\underset{|\phi\rangle\in{\mathcal{S}}}{\text{maximize}} \; \frac{\langle \phi | \hat{G}^{\dagger}\hat{\Phi}\hat{G} | \phi \rangle}{ \langle \phi| \hat{G}^{\dagger}\hat{\Sigma}\hat{G} |\phi \rangle}
\label{eq:optimization}
\end{equation}
where $\mathcal{S}$ is the set of all current sheets on a plane above the $z=0$ plane. The form of Eq.~(\ref{eq:optimization}) is a generalized eigenvalue problem involving the operators $\hat{A}:=\hat{G}^{\dagger}\hat{\Phi}\hat{G}$ and $\hat{B}:=\hat{G}^{\dagger}\hat{\Sigma}\hat{G}$. The solution is given by the two-dimensional current density that satisfies  $\hat{A}|\phi_\text{opt}\rangle=\lambda_{\text{max}}\hat{B}|\phi_\text{opt}\rangle$ where $\lambda_{\text{max}}$ is the largest generalized eigenvalue. If $\hat{B}$ is invertible, then the solution $|\phi_{\text{max}}\rangle$ can be obtained from a standard eigenvalue decomposition of the operator $\hat{B}^{-1}\hat{A}$. Numerical computation can be considerably accelerated by (i) selecting the plane wave basis, which diagonalizes the Green's function operator for the multilayer geometry, and (ii) exploiting degeneracies due to azimuthal symmetry about the focal axis. The calculation reduces to inversion of dyads at each spatial frequency, without need to explicitly form the full system matrices. This inverse filtering process is closely related to time-reversal~\cite{Tanter_2000} and can be generalized to transparent media by allowing the material loss to asymptotically approach zero.

\begin{figure}
   \centering
   \includegraphics[width=8cm]{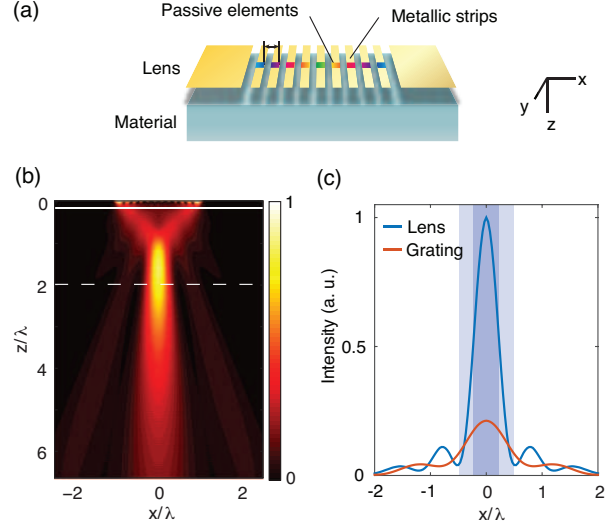} 
   \caption{Focusing across an interface with a metasurface. (a) Metasurface lens above a material with refractive index $n=2$. The strips are assumed to be periodic in the $y$ direction. (b) Magnetic field intensity in material for incident s-polarized radiation. The focal plane (dashed line) is $4\lambda/n$ from the lens. (c) Intensity profile at the focal plane for a plane wave at normal incidence. The spot width is $0.42\lambda$ FWHM for the lens (dark shading) and $\lambda$ for the grating (light shading). The grating consists of metal strips without passive elements.}
   \label{fig:focal_line}
\end{figure}

We first consider the two-dimensional geometry in Fig.~\ref{fig:focal_line} where the material has a refractive index $n=2$. For incident s-polarized radiation, we numerically solve Eq.~(\ref{eq:optimization}) to obtain a source that focuses to a line at a $4\lambda/n$ distance. A linear metasurface is used to shape a normally incident plane wave such that the exiting field matches the solution. The required impedance values of the passive elements are solved by using a point-matching method (see Appendix C). Fig.~\ref{fig:focal_line}(c) shows that line width of the focal spot is subwavelength $0.42\lambda$ FWHM. To verify that the focusing effect is due to phase (not amplitude) modulation of incident wave, the passive elements are removed such that the surface acts as a grated aperture. The focal spot for the grating is not subwavelength ($\sim\lambda$); the intensity at the focal point is also decreased by a factor of 4~[Fig.~\ref{fig:focal_line}(c)].

The physics underlying the lensing effect is substantially different from near-field focusing devices. Unlike near-field plates, which focus evanescent waves at a strictly subwavelength distance (typically less than $\lambda/10$), our lens' focusing ability results from conventional interference between propagating waves and, as a result, the focal plane can be many wavelengths away. The enhanced resolution of our lens follows from the shaping of the near-field phase profile, which couples the incident wave to forbidden angles on interaction with material, rather than the near-field interference effects of near-field plates. Because the focusing is not subject to the intrinsic decay of the near-field, the intensity at the focal spot can be comparable to or higher than the incident intensity. As with solid immersion lenses, the focusing resolution remains subject to the diffraction limit, although with spot size set by the material rather than the free-space wavelength.

\begin{figure}
   \centering
   \includegraphics[width=6.5cm]{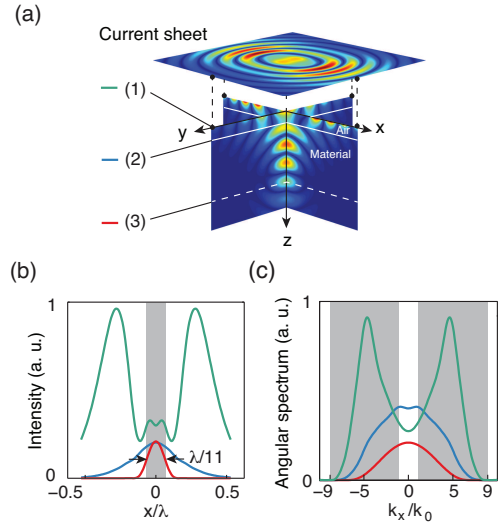} 
   \caption{Optimal focusing across an interface. (a) Focusing in material with an optimized current sheet (contour plot of $|\text{Re}(H_x)|$ shown). The material is 0.5\% saline water ($n=8.8+i0.9$ at 1.5~GHz). The focal plane (dashed white line) is located $2.3\lambda/n$ (wavelength in material) from the current sheet. (b) Magnetic field intensity profile and (c) angular spectrum at the (1) source, (2) interface, and (3) focal planes. The shaded region in (b) denotes the width of the focal spot $\lambda/11$ (FWHM). Shaded regions in (c) show the propagating wavevectors $1<|k_{x}/k_{0}|<8.8$ supported by material.}
   \label{fig:opt_focusing}
\end{figure}

Next, we consider a three-dimensional geometry where a planar source is positioned a subwavelength distance ($\lambda/15$) above a material whose refractive index at microwave frequencies approximates biological tissue (real part $n=8.8$). Due to symmetry about the focal axis, the polarization of the fields at the focal point can be arbitrarily specified. Setting the electric field to be linearly polarized in the $x$ direction, the solution to Eq.~(\ref{eq:optimization}) is found to be a surface wave consisting of concentric ring-like currents around the focal axis [Fig.~\ref{fig:opt_focusing}(a)]. In air, the resulting fields are evanescent and non-stationary, propagating in-plane towards the focal axis (see Supplemental Video). Importantly, the intensity profile at the source plane is significantly non-zero only within a finite circular region [Fig.~\ref{fig:opt_focusing}(b)]. The radius of this region defines an effective aperture size that is directly related to the loss in the material system and the depth of focus. At the focal plane in the material, the field converges to a spot of width $\lambda/11$, measured full-width half-maximum (FWHM), at a distance of about $2.3\lambda/n$ (wavelength in material) from the source plane~[Fig.~\ref{fig:opt_focusing}(b)]. Although the wave originates in air, the spot size approaches Abbe's diffraction limit $\lambda/(2 n \sin\theta_{\text{ap}})$ in homogenous material, where $\theta_{\text{ap}}$ is the half-angle the aperture subtends the focal point, due to the source's ability to access forbidden wave components [Fig.~\ref{fig:opt_focusing}(c)].

\section{Experimental Demonstration}

\begin{figure}
   \centering
   \includegraphics[width=8.5cm]{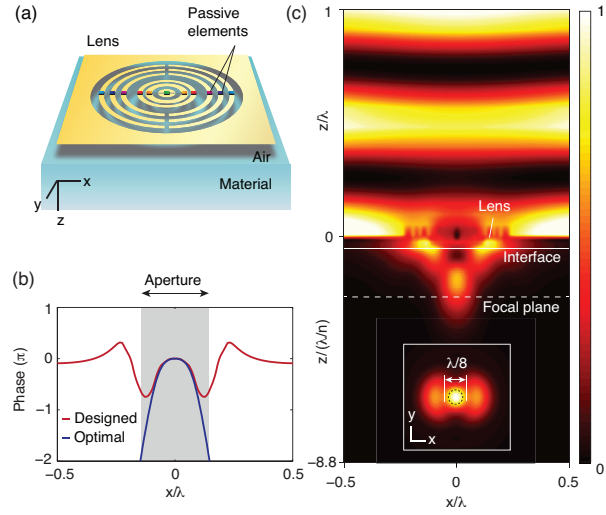} 
   \caption{Planar immersion lens with concentric rings. (a) Metasurface lens consisting of concentric rings for focusing an incident plane wave. The rings are loaded with passive elements. The lens is positioned $\lambda/15$ above saline water. (b) Phase profile of the magnetic field ($H_{x}$) exiting the lens along the $x$ axis. The phase response of the lens matches the profile dervied the optimization procedure within the aperture. (c) Contour plot of the magnetic field intensity. The focal plane is located $2.3\lambda/n$ (wavelength in material) away from the lens. Simulations are performed at 1.6~GHz.}
   \label{fig:focal_spot}
\end{figure}

Based on the solution to the optimization problem, we designed a planar immersion lens capable of focusing an incident plane wave. The lens consists of concentric rings loaded with passive elements [Fig.~\ref{fig:focal_spot}(a)] forming an aperture that spans the width of the calculated intensity distribution at the source plane. In agreement with time-reversal considerations~\cite{Tanter_2000, Lerosey_2004}, the phase response of the lens is parabolic with concavity reversed from that expected of an outward propagating wave [Fig.~\ref{fig:focal_spot}(b)]. Fig.~\ref{fig:focal_spot}(c) shows that the lens focuses to a spot of size of $\lambda/8$ FWHM. The spot size is slightly larger compared to $\lambda/11$ for the optimal current sheet due to the mismatch in the phase response outside of the finite extent of the aperture [Fig.~\ref{fig:focal_spot}(b)].

In the microwave regime, an application for the immersion lens is energy transfer through biological tissue~\cite{Ho_2014}. Since tissue exhibits a relatively large refractive indices at gigahertz frequencies (e.g., real part of $7.4$ for muscle at 1.6~GHz)~\cite{Gabriel_1996}, the efficiency of energy transfer can be substantially enhanced if the focal spot size can approach $\sim\lambda/n$. In this context, the planar nature of metasurfaces is a key advantage because it enables fabrication on flexible substrates. The currents in the rings can be set up by applying an electric dipole moment across the center element, similar to recently reported near-field plates~\cite{Imani_2013}, such that the lens can be excited through a compact electronic source rather than an incident plane wave. 

\begin{figure*}
   \centering
   \includegraphics[width=15cm]{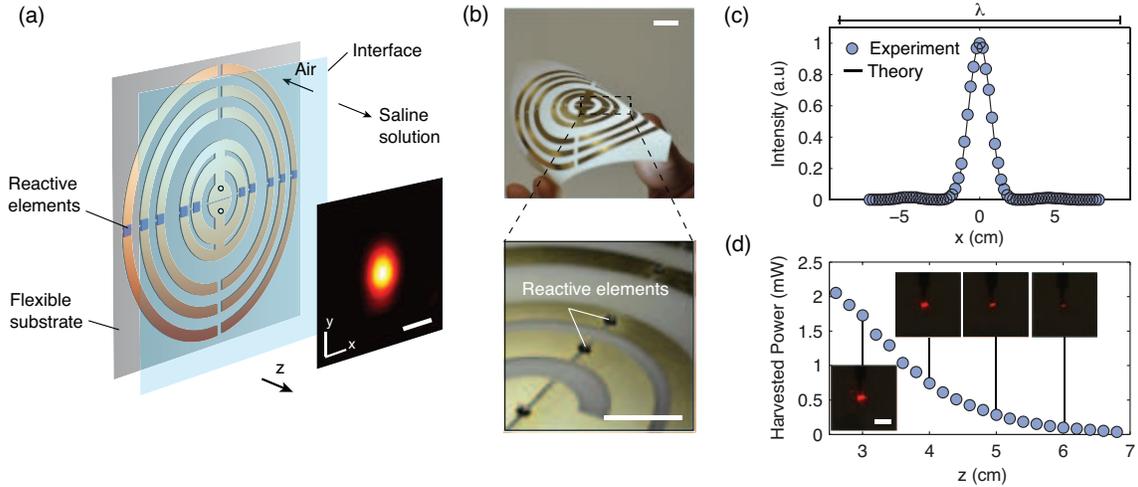} 
   \caption{Focusing in a dielectric liquid using a coaxial-excited lens. (a) Focusing in a 0.5\% saline solution with a metasurface lens. The lens is excited across the center gap (white circles) at 1.6~GHz by a coaxial cable. Separation between the lens and interface, 1~cm; focal plane and lens, 5~cm. (b) Photograph of the lens. (c) Magnetic field intensity profile at the focal spot. The theory curve is calculated using the refractive index $n=8.8+i0.9$. (d) Energy transfer to a 2-mm device as a function of distance. Inset images show the brightness of an LED on the device. The output power is 500~mW. Scale bar, 1~cm.}
   \label{fig:experiment}
\end{figure*}

We designed the lens shown in Fig.~\ref{fig:experiment}(a) consisting of modified concentric rings fabricated on an ultrathin, flexible FR4 microwave substrate (100~$\mu$m thickness) [Fig.~\ref{fig:experiment}(b)]. The reactive elements are solved assuming an electric dipole excitation across the center element. Energy transfer is demonstrated by operating the lens above a simulated tissue volume. The setup consists of a glass tank filled with saline solution with dimensions large enough to minimize reflection. We positioned the lens 1~cm above the air-solution interface and scanned a magnetic field probe through the volume. The lens couples energy into the tank with a 90\% efficiency, estimated from the scattering parameters (see Appendix A), and produces a highly symmetric focal spot 5~cm from the lens. Fig.~\ref{fig:experiment}(c) shows the magnetic field intensity profile of the focal spot. The spot diameter is about 1.7~cm (that is, $\lambda/11$) FWHM and is in close agreement with the theoretical optimal. 

Energy at the focal point can be harvested with a subwavelength coil. We attached a 1-mm radius coil on a microelectronic implant and tested the device in a liquid volume simulating biological tissue. We set the output power to 500~mW (about the power radiated by cell phones), such that radiation exposure levels are well below safety thresholds for humans~\cite{IEEE_SAR_2005}. Fig.~\ref{fig:focal_spot}(d) shows the experimental characteristics of the system. At the focal plane, the transferred power is measured to be about 290~$\mu$W. This level exceeds the power consumption of many classes of bioelectronics (a pacemaker, by comparison, consumes 8~$\mu$W). The transfer depth is approximately 50 times the radius of the coil. At this range, the transfer efficiency of the lens (about $6\times10^{-4}$) is significantly higher than standard inductive coupling systems, which are limited to efficiencies on the order of $10^{-7}$ in similar configurations~\cite{Kim_2013}. The enhanced performance is key for wireless powering to be practical, since the output power is usually limited (typically from 1 to 5~W) due to safety considerations~\cite{Ho_2014}.

\section{Conclusion}

We have demonstrated a planar device based on metasurfaces with the functionality of a solid immersion lens. The enhanced focusing resolution of the device results from the ability of metasurfaces to control the near-field with subwavelength resolution on interaction with dielectric material. At optical frequencies, planar immersion lenses could be implemented with closely-spaced plasmonic antennas~\cite{Yu_2011} or dielectric resonators~\cite{Monticone_2013}, with mutual interactions accounted for by tuning the properties of optical ``lumped'' elements~\cite{Engheta_2007,Sun_2012c}. By incorporating subwavelength structures that interact with the magnetic field component of incident radiation, the metasurface could also modify the optical impedance, allowing reflection at the interface to be eliminated~\cite{Pfeiffer_2013,Yu_2014}. As the fabrication of the metasurface is simple and planar in nature, the metasurface-based lens can be integrated into complex systems, such as nanophotonic chips or conformal biomedical devices.

\section*{Appendix A: Methods}

\setcounter{equation}{0}
\renewcommand{\theequation}{A\arabic{equation}}

\emph{Metasurface refraction.} For the refraction simulation in Fig.~2, the metasurface is modeled as a linear array of 15 copper strips in FDTD software (CST Microwave Suite). The strips are 0.5~cm wide and spaced 1~cm apart on a FR4 substrate of thickness 1.2~mm. Reflective metal sheets were placed around the aperture in order to remove edge diffraction. The structure is excited by a plane wave at varying angles of incidence at 1.5~GHz with perfect electric conductor boundaries on the $y=\pm 1$~cm planes. The predicted angles of refraction are calculated using the generalized Snell's law
\begin{equation}
\sin\theta_{\text{ref}}=\frac{1}{n}\sin\theta_{\text{inc}} + \frac{\lambda}{2\pi n}\nabla{\Phi}
\end{equation}
where $n$ is the index of refraction, $\theta_{\text{ref}}$ the angle of refraction, $\theta_{\text{inc}}$ the angle of incidence, and $\nabla\Phi$ the phase gradient~\cite{Yu_2011}. The passive components are used to create a phase gradient of $\nabla\Phi=\pi/0.55\lambda$; their values are given in the Supplemental Materials. 

\emph{Metasurface focusing.} The focusing metasurface in Fig.~\ref{fig:focal_spot} is excited by a plane wave at 1.6~GHz $y=\pm 5$~cm at normal incidence. The concentric rings are placed in a circular aperture of diameter 8.4~cm. The largest ring has a diameter of 7.5 cm. To calculate the impedance values of the passive elements, a multiport simulation of the structure is performed. The required passive elements are then obtained by solving a set of balance equations in the impedance matrix~(see Appendix C). The values of the passive elements are given in the Supplemental Materials.

\emph{Experimental setup.} The prototype of the planar immersion lens was fabricated on an ultrathin (100~$\mu$m) FR4 dielectric ($\epsilon_{r}=4.4$) substrate using commercially available surface-mount capacitors and inductors. The components are mounted on the substrate using standard soldering techniques. The lens is excited by a co-axial cable attached by a SMA connector. The connector is soldered across the center element in the lens [Fig.~5(a)], with outer and inner conductors separated by the gap across the center circular element. The prototype dimensions are 8~cm$\times$8~cm. For the experimental measurement, a liquid volume consisting of 0.5\% saline solution was used as the focusing medium. The complex permittivity of the liquid was measured to be $\epsilon_{r}=77.2 + 15.4i$ using a dielectric probe. An excitation signal (500~mW, 1.6~GHz) was sent to the lens from the signal generator (Agilent E4436B) through a power amplifier with co-axial cables. With the lens surface positioned 1~cm from the simulated tissue volume, the reflection was measured to be $S_{11}=-9.8$~dB using an Agilent E5072A VNA. Since the back-radiation (power radiated into air) is measured to be $<1\%$, the efficiency of power coupled into the solution can be calculated as $1-S_{11}\approx90\%$. The spatial distribution field in the material is measured by scanning a wideband magnetic field probe (10~MHz to 18~GHz; AEMP002, AET, Inc.) through the volume using a robotic positioner (4EM500, NEC, Inc.).

\section*{Appendix B: Degeneracy and the Matched Filter}

\setcounter{equation}{0}
\renewcommand{\theequation}{B\arabic{equation}}

The operators $\hat{A}:=\hat{G}^{\dagger}\hat{\Phi}\hat{G}$ and $\hat{B}:=\hat{G}^{\dagger}\hat{\Sigma}\hat{G}$ are symmetric under rotation about the $z$ axis and therefore commute with the transverse rotation operator $\hat{R}_{\theta}$. The generalized eigenvalue problem in Eq.~(\ref{eq:optimization}) is degenerate, since if $|\phi\rangle$ is a generalized eigenvector, then any rotation $\hat{R}_{\theta}|\phi\rangle$ is also a generalized eigenvector. At the focal point, $\hat{R}_{\theta}$ rotates the transverse polarization of the field vector. The right-handed (RHP) and left-handed (LHP) circular polarization states are the eigenvectors of $\hat{R}_{\theta}$ and span the space of solutions. Because any transverse polarization can be generated by the superposition of RHP and LHP states, the $x$ and $y$ components of the polarization at the focal point can be arbitrarily fixed without loss of optimality. In this case, the generalized eigenvalue problem reduces to a matched filtering problem. To see this, let $\hat{\mathbf{n}}$ be a unit vector in the direction of the polarizability vector $\alpha\hat{\mathbf{n}}$. The polarization current at the focal point is then
\begin{equation}
\mathbf{j}_{p}=\frac{\partial\mathbf{P}}{\partial t}=\alpha\delta(\mathbf{r}_{f}) \left( \frac{\partial \mathbf{E}}{\partial t}\cdot\hat{\mathbf{n}}\right)\hat{\mathbf{n}}
\end{equation}
Adopting phasor notation with time dependency $\exp(-i\omega t)$, the rate of work performed at the focal point is
\begin{equation}
\frac{1}{2}\text{Re}\left[ \int d\mathbf{r}~\mathbf{j}_{p}^{*}\cdot\mathbf{E}\right]=\frac{1}{2}\omega\alpha'' |\mathbf{E}(\mathbf{r}_{f})\cdot\hat{\mathbf{n}}|^2
\label{eq:eff_polarized}
\end{equation}
which yields the following expression for efficiency
\begin{equation}
\eta = \frac{\alpha'' | \mathbf{E}(\mathbf{r}_f)\cdot \hat{\mathbf{n}}|^2}{ \int d\mathbf{r}~\epsilon''(\mathbf{r})~|\mathbf{E}|^2}
\end{equation}
Let $\hat{\Phi}'=\int d\mathbf{r}~\alpha'' \delta(\mathbf{r}-\mathbf{r}_{f},\mathbf{r}'-\mathbf{r}_{f}) \hat{\mathbf{n}}\hat{\mathbf{n}}$ be the new focal point selection operator, where $\hat{\mathbf{n}}\hat{\mathbf{n}}$ is a dyad. Because of the polarization vector $\hat{\mathbf{n}}$, $\hat{\Phi}'$ can now be written as an outer product $|\chi\rangle\langle\chi|$. In operator notation, Eq.~(\ref{eq:eff_polarized}) becomes
\begin{equation}
\eta = \frac{\langle \phi | \hat{G}^{\dagger}\hat{\Phi}\hat{G} | \phi \rangle}{ \langle \phi| \hat{G}^{\dagger}\hat{\Sigma}\hat{G} |\phi \rangle } = \frac{|\langle\phi| \hat{G} |\chi \rangle|^{2}}{ \langle \phi| \hat{G}^{\dagger}\hat{\Sigma}\hat{G} |\phi \rangle}
\end{equation}
which can be maximized by the matched filter. It can be shown from the Cauchy-Schwarz inequality~\cite{Kim_2013} that $\eta$ is maximized when
\begin{equation}
|\phi_{\text{opt}}\rangle = (\hat{G}^{\dagger}\hat{\Sigma}\hat{G})^{-1} \hat{G} |\chi\rangle
\label{eq:matched_filter_solution}
\end{equation}
For a multilayer material system, $\hat{G}$ is diagonal for the angular spectrum basis. Evaluation of Eq.~(\ref{eq:matched_filter_solution}) requires only the inversion of individual dyads at each $(k_x,k_y)$ component, without need to form the full system matrices.

\section*{Appendix C: Selection of Impedance Elements}

\setcounter{equation}{0}
\renewcommand{\theequation}{C\arabic{equation}}

We consider a multiport structure ($N$ ports) composed of primitive elements designed to induce a desired field pattern at the air-material interface. The structure of the elements are selected according to the dominant mode of interaction (e.g. rings for magnetic response or strips for electric response) and are spaced with an appropriate sampling period to discretize the output field. Let $F_{0}$ be a Cartesian component of the desired field, $z$ the direction of focus, and $z=z_{0}$ the interface plane. For simplicity, consider the two-dimensional case, with the geometry being translationally invariant along the $y$ axis. We begin by applying a point-matching method, similar to the design procedure for near-field plates~\cite{Grbic_2011}, to solve for the desired amplitudes and phases $a\in\mathbb{C}^{N}$ at each port. To do so, a set of fields $F_{n}$, corresponding to the field pattern when the $n$th port is excited, is first obtained with a field solver. We then construct vectors $f_{0}$ and $f_{1}, \ldots, f_{N}$ by sampling the fields along the interface. From these vectors, we define a transfer matrix $T= [f_{1}, f_{2}, \ldots, f_{N}]$. For a set of weights $a$ at the ports, the total field at the interface can be found as $f=Ta$. The optimal weights are found by solving a least squares problem minimizing the error between $f$ and $f_{0}$, which has solution $a_{\text{opt}}=(T^{H}T)^{-1}T^{H} f_{0}$.

The goal is now to find a set of passive impedances $\{ Z_{R,n} : \text{Re}(Z_{R,n})\geq0, n=1,\ldots,N \}$ to terminate all or a subset of the ports such that port amplitudes respond with $a_{\text{opt}}$ under a specific excitation (plane wave or single port). To determine the values of these elements, we first characterize the coupling between ports with the mutual impedance matrix $Z\in\mathbb{C}^{N\times N}$. We then find the vector of voltages $v_{S}$ at each port under excitation. For a plane wave excitation, these are the voltages measured at each port under open-circuit conditions, while for an excitation at Port 1, $v_{S}$ is the vector $v_{S}= [v_{0}, 0, \ldots, 0]^{T}$, where $v_{0}$ is the applied excitation voltage. These voltages satisfy the balance equations
\begin{align}
v_{S}+v_{I}&=Z i_{C} \\
v_{I}&=-Z_{R} i_{C}
\end{align}
where $i_{C}$ is vector of currents and $Z_{R}$ is a diagonal matrix with diagonal entries $Z_{R,n}$. To find the impedances $Z_{R,n}$, we rearrange the above to obtain
\begin{equation}
v_{I}=-\left(I+Z_{R}Z^{-1}\right) ^{-1}Z_{R}Z^{-1}v_{S}
\end{equation}
where $I$ is the identity matrix. Defining $a:=v_{I}/v_0$ and $e_{0}:=v_{S}/v_0$ and rearranging, we find
\begin{equation}
Z_{R}(Z^{-1} \left( e_{0}+a)\right ) = -a
\end{equation}
To force the port amplitudes to be $a_{\text{opt}}$, the passive elements are thus required to have the impedance values
\begin{equation}
Z_{R,n}= -a_{\text{opt},n} / \zeta_{n}
\end{equation}
where $\zeta_{n}$ are the entries of the vector $\zeta:=Z^{-1}\left(e_{0}+a_{\text{opt}}\right)$.

\clearpage

\end{document}